\documentclass[twocolumn]{IEEEtran}
\usepackage[T1]{fontenc}
\usepackage[latin9]{inputenc}
\usepackage{color}
\usepackage[table,svgnames]{xcolor} 
\usepackage{array}
\usepackage{enumitem}
\usepackage{graphicx}
\usepackage{tabularx}
\usepackage{makecell}
\usepackage{booktabs}
\usepackage{times}
\usepackage[unicode=true,
 bookmarks=true,bookmarksnumbered=true,bookmarksopen=true,bookmarksopenlevel=1,
 breaklinks=false,pdfborder={0 0 0},pdfborderstyle={},backref=false,colorlinks=false]
 {hyperref}
\hypersetup{pdftitle={Your Title},
 pdfauthor={Your Name},
 pdfpagelayout=OneColumn,pdfnewwindow=true, pdfstartview=XYZ, plainpages=false}

\makeatletter

\usepackage[caption=false,font=footnotesize]{subfig}
\usepackage{algorithm}
\usepackage{algorithmic}

\usepackage{multirow} %multirow for format of table 
\usepackage{amsmath} 
\usepackage{xcolor}

\allowdisplaybreaks[4]

\ifCLASSOPTIONcompsoc
\usepackage[caption=false,font=normalsize,labelfont=sf,textfont=sf]{subfig}
\else
\usepackage[caption=false,font=footnotesize]{subfig}
\fi

\usepackage{cite}
\usepackage{bm}
\usepackage{algorithmic}
\usepackage{algorithm}
\usepackage{graphicx}
\IEEEoverridecommandlockouts

\usepackage{lettrine}

\makeatother

\begin{document}
\title{\textcolor{black}{A Blockchain-Based Trust Framework for Resilient Cross-Domain UAV Service Orchestration}\textcolor{blue}{\index{}}}
\author{Yao Wu, Ziye Jia, \IEEEmembership{Member,~IEEE}, Jingjing Zhao, \IEEEmembership{Senior Member,~IEEE}, Haoyang Wang,\\ Qihui Wu, \IEEEmembership{Fellow,~IEEE}, and Zhu Han, \IEEEmembership{Fellow,~IEEE}

\thanks{This work was supported in part by National Natural Science Foundation of China under Grant 62231015, in part by the Open Project Program of State Key Laboratory of CNS{/}ATM (No.2025B11), in part by the Jiangsu Provincial Key Research and Development Program under Grant BE2022068 and under Grant BE2022068-1,  in part by the Young Elite Scientists Sponsorship Program by CAST 2023QNRC001, and in part by the Postgraduate Research \& Practice Innovation Program of Jiangsu Province under Grant SJCX25\_0152. (\textit{Corresponding author: Ziye Jia}).}
\thanks{
Yao Wu, Ziye Jia, Haoyang Wang and Qihui Wu are with the College of Electronic
and Information Engineering, Nanjing University of Aeronautics and
Astronautics, Nanjing 211106, China, (e-mail: wu\_yao@nuaa.edu.cn, jiaziye@nuaa.edu.cn, wanghaoyang@nuaa.edu.cn, wuqihui@nuaa.edu.cn).

Jingjing Zhao is with the School of Electronics and Information En- gineering, Beihang University, 100191, Beijing, China, and also with the State Key Laboratory of CNS/ATM, 100191, Beijing, China, (e-mail:jingjingzhao@buaa.edu.cn). 

%Bomin Mao is with the School of Cybersecurity, Northwestern Polytechnical University, Xi'an 710129, China, (e-mail:maobomin@nwpu.edu.cn).

Zhu Han is with the University of Houston, Houston, TX 77004 USA, and also with the Department of Computer Science and Engineering, Kyung Hee University, Seoul 446-701, South Korea, (e-mail: hanzhu22@gmail.com).

\textit{\textcolor{black}{}}}}
\maketitle
\begin{abstract}
Unmanned aerial vehicle (UAV) networks are increasingly deployed for complex missions, including disaster response, intelligent logistics, and environmental monitoring. These missions generally  require coordinated collaboration among multiple UAVs across distinct administrative domains. To support such cross-domain cooperation, service function chains (SFCs) are constructed, where complex workflows are decomposed into ordered service functions assigned to appropriate UAVs along the mission path. However, it is challenging to ensure secure, trustworthy, and low-latency cross-domain SFC orchestration in identity management, authentication, and resilience to node failures. To address these issues, this paper proposes a consortium blockchain-based trust architecture for cross-domain decentralized identity verification, auditable task execution, and dynamic service-aware orchestrator selection. The framework employs a hierarchical four-phase cross-domain authentication protocol covering the credential pre-verification, intra-domain execution, secure relay, and audit logging. The  use case analysis  confirms that the proposed framework achieves substantial reductions in authentication latency and significant improvements in system throughput against centralized and static schemes. The open challenges in scalability, adaptive trust assessment, interoperability, and energy efficiency are discussed, thereby providing directions for future researches on secure and efficient cross-domain UAV service orchestration.
\end{abstract}

\begin{IEEEkeywords}
Unmanned aerial vehicle (UAV), service function chain (SFC), blockchain, trust framework.
\end{IEEEkeywords}

\section{Introduction}
Unmanned aerial vehicle (UAV) networks are rapidly evolving from isolated single-unit deployments into coordinated multi-agent systems that provide significant advantages in disaster response, intelligent logistics, and precision agriculture~\cite{zeng2016wireless, yan2025multi}. The complex missions frequently surpass the operational limits of a single UAV functioning alone in sensing, computation, communication, and endurance, requiring collaborations among multiple heterogeneous platforms across distinct administrative domains~\cite{lu2024uav}. The service function chain (SFC) establishes a structured approach for these cooperative missions by decomposing complex tasks into ordered sequences of functional components including sensing, data processing, and relaying, with each component assigned to an appropriate node~\cite{wu2024adaptive}. By mapping sensing, processing, and relaying functions onto different UAVs in an ordered manner, SFCs integrate distributed sensing and computation into an end-to-end mission workflow, thereby improving mission flexibility and operational efficiency~\cite{wei2024hierarchical}.

However, the secure and reliable SFC execution in dynamic open airspace introduces a distinct set of challenges that stem from the rapid spatial mobility of UAVs and the absence of stable trust anchors \cite{yang2023toward}. The continuous topology variation degrades the persistence of trust relationships, and the short-lived communication sessions make timely credential validation more difficult. These factors complicate the establishment of reliable end-to-end trust and create conditions where conventional credential management fails to ensure uninterrupted SFC execution. The exposed inter-UAV communication channels are susceptible to eavesdropping, message tampering, and impersonation attacks. The frequent mobility and dynamic joining or departure of nodes further complicate the authentication, access control, and behavior verification~\cite{tan2022blockchain}. The traditional centralized authentication frameworks, which typically rely on the trusted authority (TA) for credential management, provide administrative simplicity but exhibit limited scalability, slow responsiveness to frequent topology changes, and an inherent single point of failure~\cite{hafeez2023blockchain}. As the UAV networks scale, these limitations increasingly undermine the mission resilience and continuity.

\begin{figure*}[t]
\centering
\includegraphics[scale=0.7]{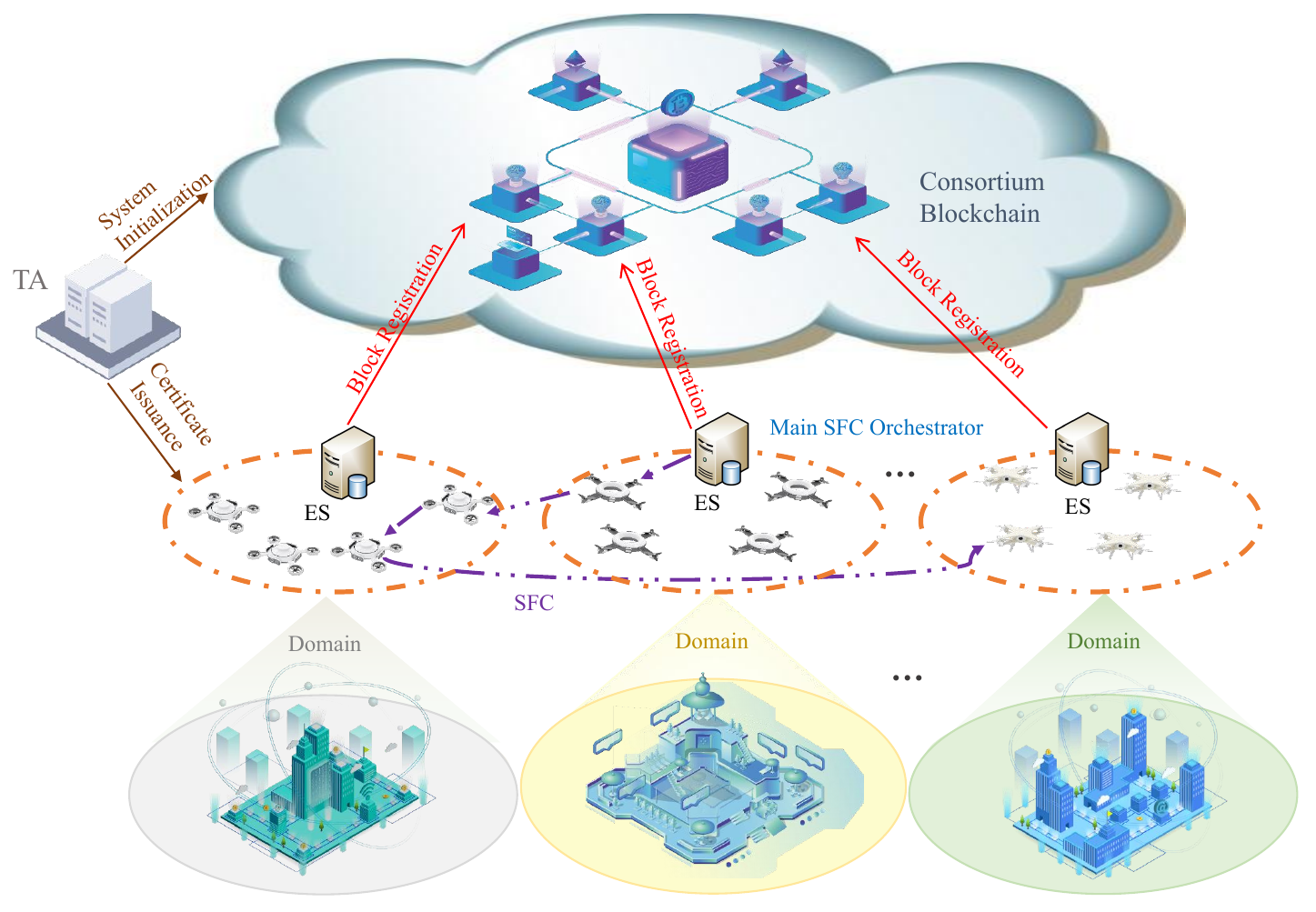}
\caption{Architecture of the proposed consortium blockchain-enabled cross-domain UAV SFC trust framework.}
\label{fig:architecture}
\end{figure*}

The challenge becomes more intricate when an SFC traverses multiple autonomous domains since the trust establishment must span administrative boundaries. The lack of shared credential governance forces each domain to verify external identities and authorize tasks under heterogeneous policies. It increases the coordination overhead and necessitates a mechanism  able to maintain the trust continuity during the domain transitions. In the cross-domain orchestration, three interrelated \emph{challenges} emerge. \emph{First}, the absence of a shared, and decentralized trust foundation hinders reliable credential transfer and identity verification across administrative boundaries~\cite{chen2022xauth}. \emph{Second}, the static or pre-assigned orchestrators fail to adapt effectively to evolving mission contexts or network topologies, resulting in coordination inefficiencies and vulnerability to orchestrator failures~\cite{gong2023lcdma}. \emph{Third}, the existing cross-domain authentication procedures often rely on the repeated global verification, which elevates the communication latency and jeopardizes real-time service continuity~\cite{pu2024litegap}. Thus, an effective solution must integrate distributed trust establishment, adaptive orchestration, and secure, low-latency inter-domain authentication within a unified framework~\cite{tan2022}.

To address these challenges, this paper proposes a blockchain-anchored trust framework for resilient cross-domain UAV SFC orchestration. The proposed architecture leverages a consortium blockchain as a decentralized trust anchor, enabling immutable recording of credential events and auditable verification without dependence on a central authority~\cite{mao2025blockchain}. On this basis, a lightweight, service-aware orchestrator election algorithm dynamically selects optimal coordinators based on the mission context and resource availability, thereby enhancing the adaptability and fault tolerance. Additionally, a hierarchical four-stage authentication and handover protocol is designed to facilitate the rapid and secure inter-domain credential migration while preserving the end-to-end integrity and verifiability. These components constitute an integrated architecture that unifies the distributed trust management, adaptive orchestration, and efficient cross-domain authentication for mission-critical UAV operations. In short, the key contributions of this work are summarized as follows.

\begin{itemize}
\item We propose a blockchain-anchored trust framework that establishes a consortium blockchain as a decentralized root of trust, enabling the tamper-evident credential management and verifiable SFC coordination across multiple domains.
\item We design a service-aware orchestrator election algorithm that dynamically selects orchestrators according to real-time mission requirements and network conditions, effectively mitigating the single-node bottlenecks and improving the orchestration efficiency.
\item We develop a four-stage cross-domain authentication and handover protocol that secures the inter-domain service transitions, achieving low-latency credential transfer and comprehensive post-mission accountability through immutable on-chain records.
\end{itemize}

\section{Blockchain-Enabled Trust Framework: Architecture Overview}

The proposed framework leverages a consortium blockchain as an immutable and distributed trust anchor. Built on this foundation, a dynamic orchestration algorithm and a lightweight cross-domain authentication protocol are jointly designed. By replacing the vulnerable centralized control with decentralized collaboration, the framework ensures reliable trust transfer and efficient coordination across administrative boundaries. As shown in Fig.~\ref{fig:architecture}, the architecture comprises four primary entities that collaborate to maintain the global trust and ensure the secure inter-domain service execution.

\begin{figure*}[t]
\centering
\includegraphics[scale=0.4]{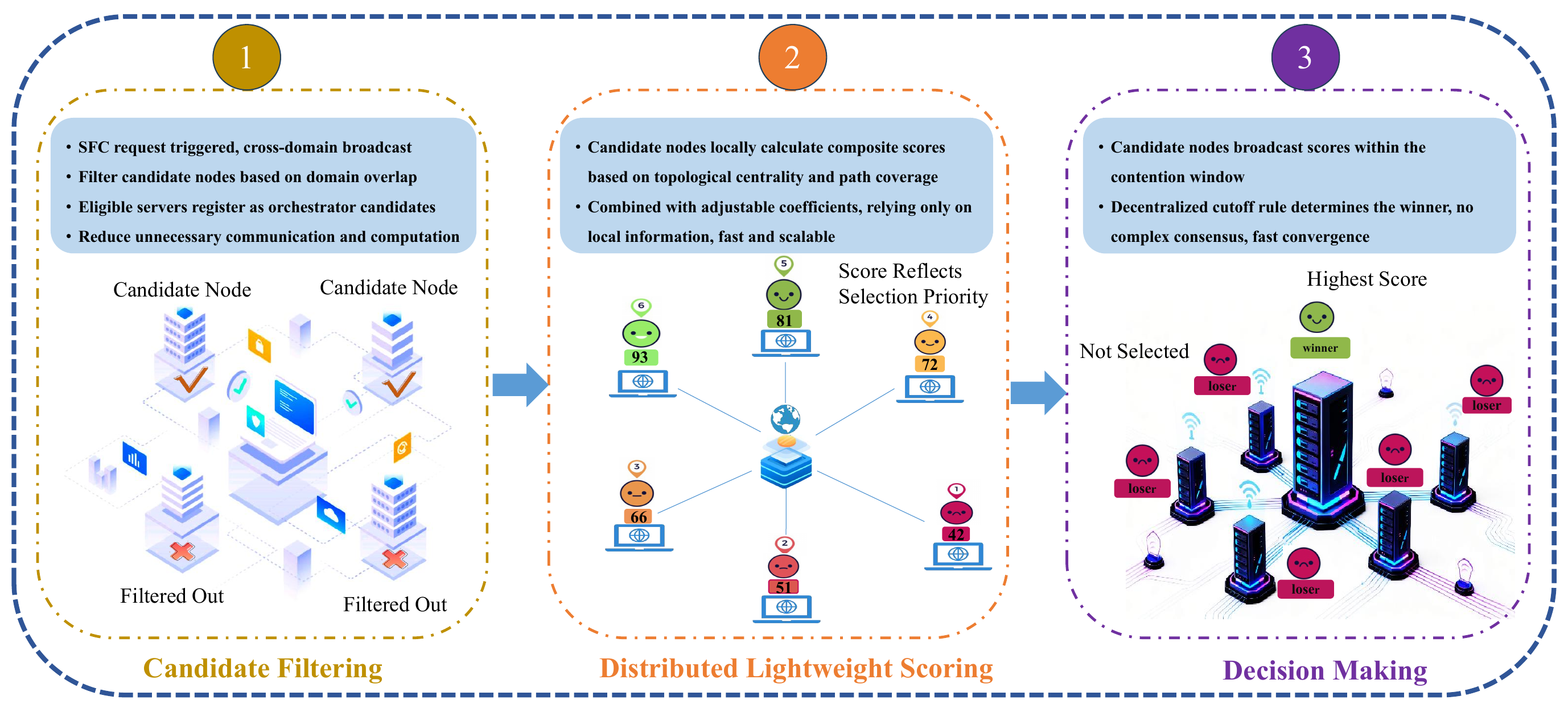}
\caption{Workflow of the proposed dynamic and service-aware orchestrator election algorithm for cross-domain UAV SFC.}
\label{fig:election}
\end{figure*}

\subsection{Core Entities}

\subsubsection{UAV}
The UAVs serve as service execution nodes within the framework. Each UAV is equipped with lightweight virtualized network function modules that support mission-specific operations such as traffic filtering, packet forwarding, data compression, format conversion, encryption and decryption, and onboard inference for target detection. After receiving the  service credentials, a UAV verifies the authenticity, executes its assigned function, and securely forwards the corresponding authorization token to the next node in the SFC chain. Each UAV is provisioned with a globally unique identifier and hardware-protected cryptographic keys to ensure the authenticity, traceability, and non-repudiation.

\subsubsection{Edge Server (ES)}
Each ES operates as the domain controller in its administrative region and undertakes two principal responsibilities. It manages the UAV identity registration, operational-state maintenance, and revocation within its domain. It also serves as a consortium blockchain node participating in Practical Byzantine Fault Tolerance consensus and maintaining the distributed ledger that stores identity data and operational logs. During the SFC execution, the ES validates the inter-domain credentials and issues the authorization certificates to UAVs in its domain, thereby supporting the secure trust propagation across multiple administrative boundaries.

\subsubsection{Consortium Blockchain}
The authorized ES nodes collaboratively maintain the consortium blockchain, which serves as the underlying trust foundation. It provides a semi-decentralized verifiable environment for storing UAV identities, public keys, and mission-related events such as service initiation, completion timestamps, and cross-domain handover logs. By replacing traditional centralized databases, the blockchain provides tamper-resistant storage and supports peer-to-peer trust establishment. The trust queries and identity verification can be performed directly on-chain without reliance on a central authority.

\subsubsection{TA}
The TA serves as an offline and strictly controlled root of trust responsible for the system initialization. It generates global cryptographic parameters and grants root certificates that authorize ESs within each domain. After the initialization, the TA remains offline while the domain enrollment, UAV registration, and credential management are executed collaboratively by the consortium blockchain and ESs. This design follows the principle of centralized initialization and decentralized operation to minimize the attack surface of the core trust authority.

\subsection{Framework Workflow}

As illustrated in Fig.~\ref{fig:architecture}, the workflow begins when a mission requester such as a ground control station issues an SFC task request to the ES cluster. A dynamic election algorithm selects an appropriate orchestrator from the ES cluster. The orchestrator queries the consortium blockchain to verify the legitimacy of participating UAVs and then issues a secure authorization credential (SAC). During the task execution, the SAC serves as a digital service token that is propagated among UAVs across multiple domains. Its validity is verified through blockchain-based signatures and public-key cryptography. In this manner, a continuous and verifiable chain of trust is established throughout the mission.

The distributed ledger eliminates the single-point-of-failure risk inherent in centralized authentication. The dynamic orchestrator election enhances the scalability, load balancing, and fault tolerance. The lightweight cross-domain authentication protocol enables low-latency service handover in dynamic environments. These capabilities provide a secure, efficient, and trustworthy foundation for the large-scale cross-domain UAV collaboration.

\section{Dynamic and Service-Aware Orchestrator Election}

The orchestrator serves as the core control entity responsible for initiating and coordinating each cross-domain SFC task. The  conventional architectures typically designate a fixed high-performance ES as a permanent orchestrator \cite{zhou2019online}. While this configuration simplifies the management overhead, it poses several notable limitations. First, it creates a single point of failure that jeopardizes the reliability of the entire workflow. If the orchestrator becomes unavailable due to network disruption or hardware malfunction, the entire SFC operation is interrupted. Moreover, a static orchestrator cannot adapt to variations in the network topology or mission requirements \cite{dolati2023layer}, resulting in inconsistent service latency across domains.

To overcome these constraints, a dynamic and service-aware orchestrator election algorithm is developed. The algorithm allocates orchestration responsibilities to the most suitable node to enhance the system resilience and operational efficiency. The principal concept is to replace traditional hardware-based selection criteria that depends on the computational capacity or memory resources with task-oriented evaluation that reflects the characteristics of each SFC instance. The selected orchestrator must demonstrate sufficient computational capability and advantageous topological placement. It is expected to be located near most participating UAVs and to minimize the cost of inter-domain authentication.
Fig.~\ref{fig:election} illustrates the lightweight election process that is automatically triggered when a new SFC request is generated. The procedure comprises \emph{three operational phases}.

\subsubsection{Candidate Filtering}

Upon detecting or receiving an SFC request, the ES broadcasts the request to other domain servers rather than processing it immediately. The request includes administrative domain metadata consisting of the domain identifiers of all domains traversed by the planned SFC. Each receiving server conducts a local evaluation by checking whether its managed domain identifier appears in this identifier list. Only servers with matched domain identifiers register as candidates. This initial filtering step ensures that only relevant nodes participate in the election, thereby minimizing the unnecessary communication and computational overhead.

\subsubsection{Distributed Lightweight Scoring}

Each candidate independently evaluates its suitability by calculating a composite score based on two performance metrics, topological centrality and path coverage. The topological centrality represents the relative position of the server within the network. The preference is assigned to candidates that act as communication hubs. Moreover, the topological centrality metric incorporates the operational weights of domains along the SFC path, so that domains handling heavier service loads contribute proportionally. This design ensures that the servers located in network-efficient positions are prioritized.  The path coverage measures the proportion of UAVs associated with the task  managed by a given candidate. A higher value indicates that the server can directly coordinate more UAVs involved in the operation, thereby reducing the inter-domain control overhead. The final score is computed by combining the two metrics with adjustable weighting coefficients, allowing flexible optimization according to specific mission objectives.  These coefficients are set according to mission priorities and network conditions, where latency-sensitive missions assign higher weight to topological centrality, and coordination-sensitive missions assign higher weight to path coverage to reduce inter-domain control overhead. The computation depends exclusively on the local topology information and task parameters and does not require real-time global synchronization. This feature enables fast and scalable execution in dynamic environments.

\subsubsection{Decision Making}

After completing the scoring process, all candidates enter a short decision interval. During this defined time window, each server broadcasts its score to other participants. A decentralized cutoff rule determines the outcome. A server declares itself as the elected orchestrator if it receives no higher score from peers during the interval. If a higher score is detected, the server withdraws from contention and acknowledges the superior candidate as the orchestrator. This method avoids multi-round message exchanges or complex consensus protocols. It enables rapid convergence and is particularly effective for dynamic and time-sensitive edge environments.

\begin{figure}[t]
\centering
\includegraphics[scale=0.3]{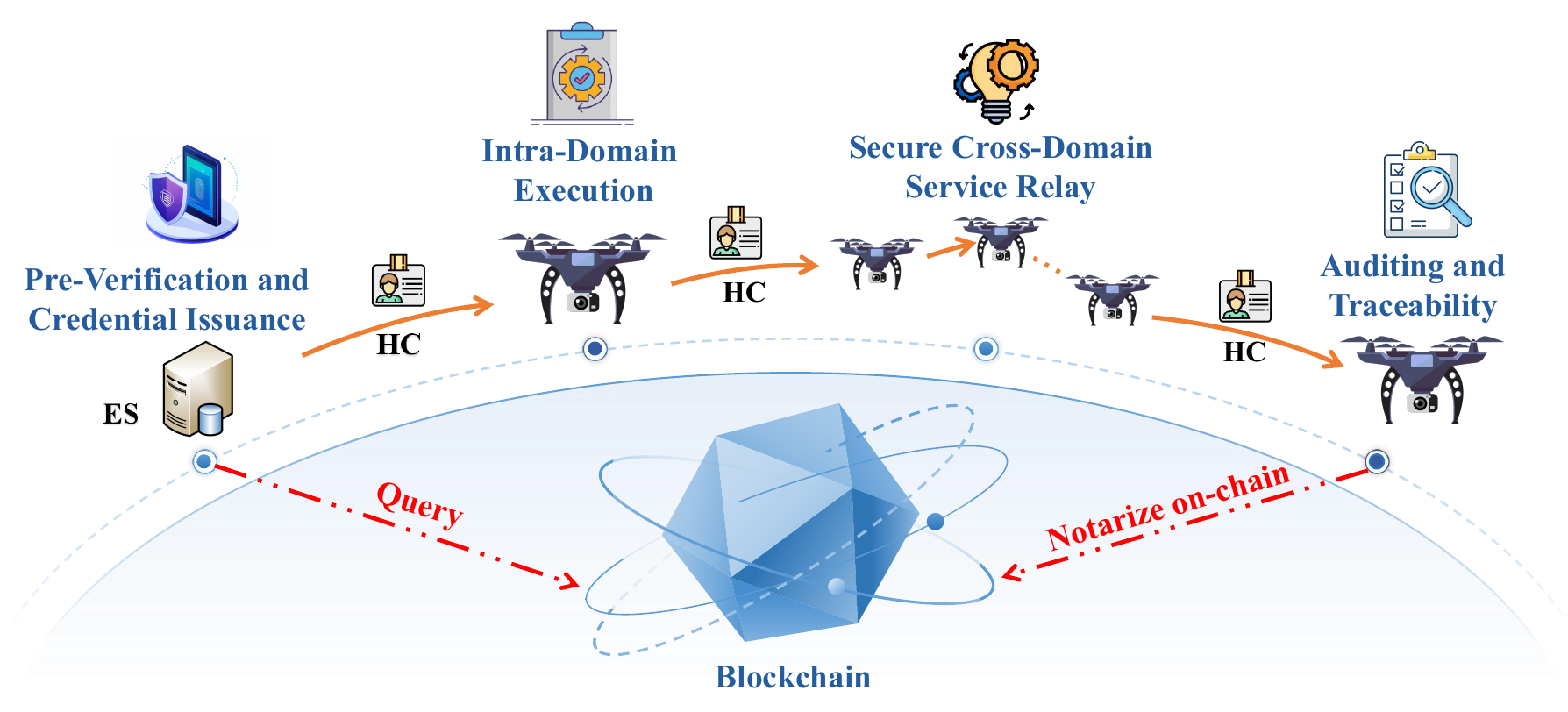}
\caption{Four-phase cross-domain SFC authentication protocol.}
\label{fig:auth_flow}
\end{figure}

\section{Cross-Domain Authentication Process}

After dynamically electing the optimal orchestrator, the system proceeds to the core stage of secure SFC execution, which involves cross-domain authentication. Achieving a reliable end-to-end trust chain in open wireless environments, without compromising continuity or latency, presents a substantial technical challenge. To address these issues, a hierarchical four-phase authentication protocol is designed. This protocol combines a one-time global verification with localized credential handovers, achieving a balance between the security and operational efficiency. As illustrated in Fig.~\ref{fig:auth_flow}, the protocol consists of four stages: pre-verification and credential issuance, secure initialization, secure cross-domain relay, and auditing and traceability.

\subsubsection{Pre-Verification and Credential Issuance}

The objective of this phase is to perform a system-wide security verification before initiating the SFC task and to issue a global access credential that eliminates delays caused by sequential authentication. The elected orchestrator queries the consortium blockchain to confirm that all UAVs along the planned SFC path are properly registered and authorized for the mission. The query is executed atomically, meaning that the verification of all UAVs along the planned SFC path is treated as an indivisible operation that either succeeds completely or fails completely. Upon successful verification, the orchestrator generates an SAC, which serves as a cryptographically protected and time-limited mission token containing essential metadata, including the task identifier and validity period. The full SFC path is encrypted using the public key of the first UAV to ensure that only the intended recipient can access the path information. The orchestrator signs the SAC with its private key to guarantee authenticity and prevent tampering. The signed SAC is then transmitted to the first UAV in the chain.

\subsubsection{Intra-Domain Execution}

Upon receiving the SAC, the first UAV initiates local verification. The UAV verifies the digital signature issued by the orchestrator to ensure the integrity and authenticity of the received credential. It then decrypts the encrypted portion of the SAC using its private key to retrieve the full SFC path and verify its role as the initiating node. The UAV also checks the validity period of the credential to prevent replay attacks. Once the verification is completed, the UAV is granted execution authorization and begins performing its assigned virtual network function, such as data filtering or format conversion. This step securely launches the SFC task under the cryptographic control.

\subsubsection{Secure Cross-Domain Service Relay}

This phase establishes the core of cross-domain trust propagation. After completing its local virtual network function, a UAV transfers the task securely to the subsequent node, which may belong to a different administrative domain. To achieve this, the UAV generates a handover credential (HC), which is a lightweight digital token containing the task identifier, the identity of the subsequent recipient, and a fresh timestamp. The HC is protected by two layers of cryptography. First, it is digitally signed by the edge server of the sender domain using the domain's private key, providing the domain-level attestation. Second, it is encrypted with the public key of the recipient UAV to ensure confidentiality and correct reception. The signed and encrypted HC is transmitted to the target UAV over an open channel. Upon receiving the HC, the target UAV verifies the digital signature from the sender domain, decrypts the HC using its private key, and confirms both identity and credential freshness. If the verification is successful, the UAV obtains execution authorization, performs the designated virtual network function, and generates a new HC for the next node. This iterative process establishes a secure trust chain that extends across multiple domains until the task completion.

\subsubsection{Auditing and Traceability}

After the final UAV in the SFC path completes its assigned function, the system enters the auditing phase. To ensure the  accountability, failure analysis, and post-mission verification, the participating domains submit essential handover logs to the Consortium Blockchain. These logs typically include immutable records such as task identifiers, node identities, timestamps, and operation types. The blockchain maintains a verifiable and tamper-evident record of the entire SFC execution lifecycle. The authorized regulators or domain administrators can query the ledger to reconstruct mission progress, enabling transparent supervision and forensic verification. This immutable audit layer enhances the integrity and credibility of the overall system.

\begin{figure}[t]
    \centering
    \includegraphics[scale=0.3]{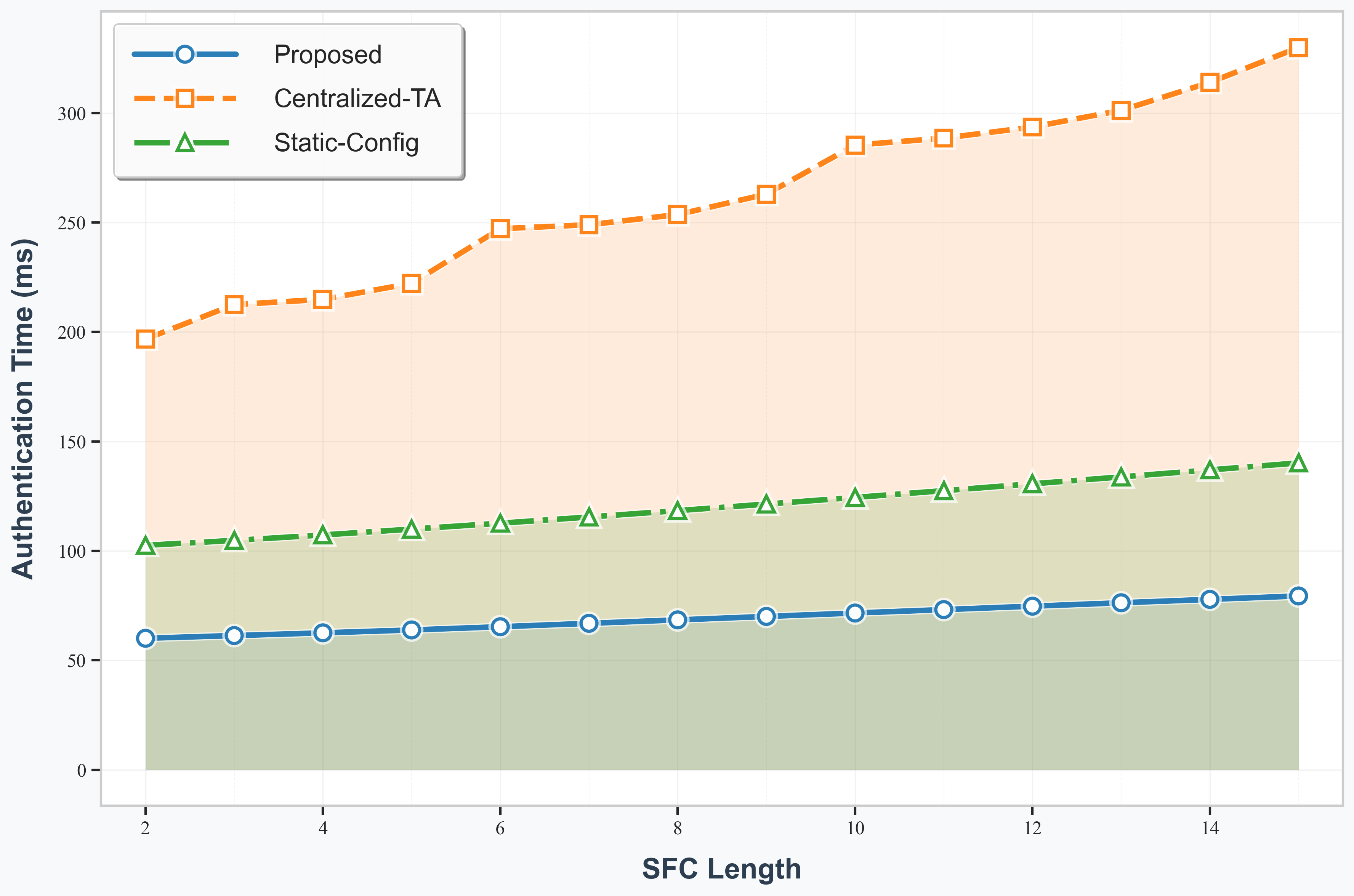}
    \caption{Authentication latency comparison under varying SFC length.}
    \label{fig:auth_latency}
\end{figure}

\section{Case Studies}

To evaluate the effectiveness of the proposed blockchain-enabled trust framework in practical UAV SFC scenarios, we conduct a comparative study with two representative baselines. The evaluation considers two critical performance indicators for cross-domain UAV operations: authentication latency and system throughput. The compared schemes include the centralized-TA method, which depends on a single trusted authority for all authentication activities, the static-config method, which employs a fixed orchestrator without dynamic selection, and the proposed method, which introduces a blockchain-enabled dynamic trust algorithm.

\subsection{Authentication Latency Analysis}

The authentication latency refers to the elapsed time from the initiation of an SFC task to the point when all participating nodes are successfully verified. This metric directly affects service responsiveness and mission continuity. Fig.~\ref{fig:auth_latency} illustrates the latency patterns of the three schemes as the SFC path length increases. All methods show longer latency as the SFC path length increases. The centralized-TA method exhibits the highest latency, because each UAV is required to interact repeatedly with a remote trust authority. This design leads to substantial cross-domain communication delay. The static-config method provides moderate improvement by removing repeated trust authority interactions. However, its fixed orchestrator is unable to adapt to path variations, and latency still increases rapidly for extended chains. In contrast, the proposed method demonstrates consistently lower and more stable latency. This advantage comes from two mechanisms. First, the pre-verification process enables the orchestrator to complete a one-time global validation of all UAV entities through a single blockchain query, which removes the need for sequential verification. Second, the cross-domain task relay relies on lightweight HC. Most validation is completed within or near each domain, which reduces the cross-domain signalling overhead. The experimental data confirm that the proposed method meets strict latency requirements for long-chain and large-scale SFC operations.

\subsection{System Throughput Analysis}

\begin{figure}[t]
    \centering
    \includegraphics[scale=0.3]{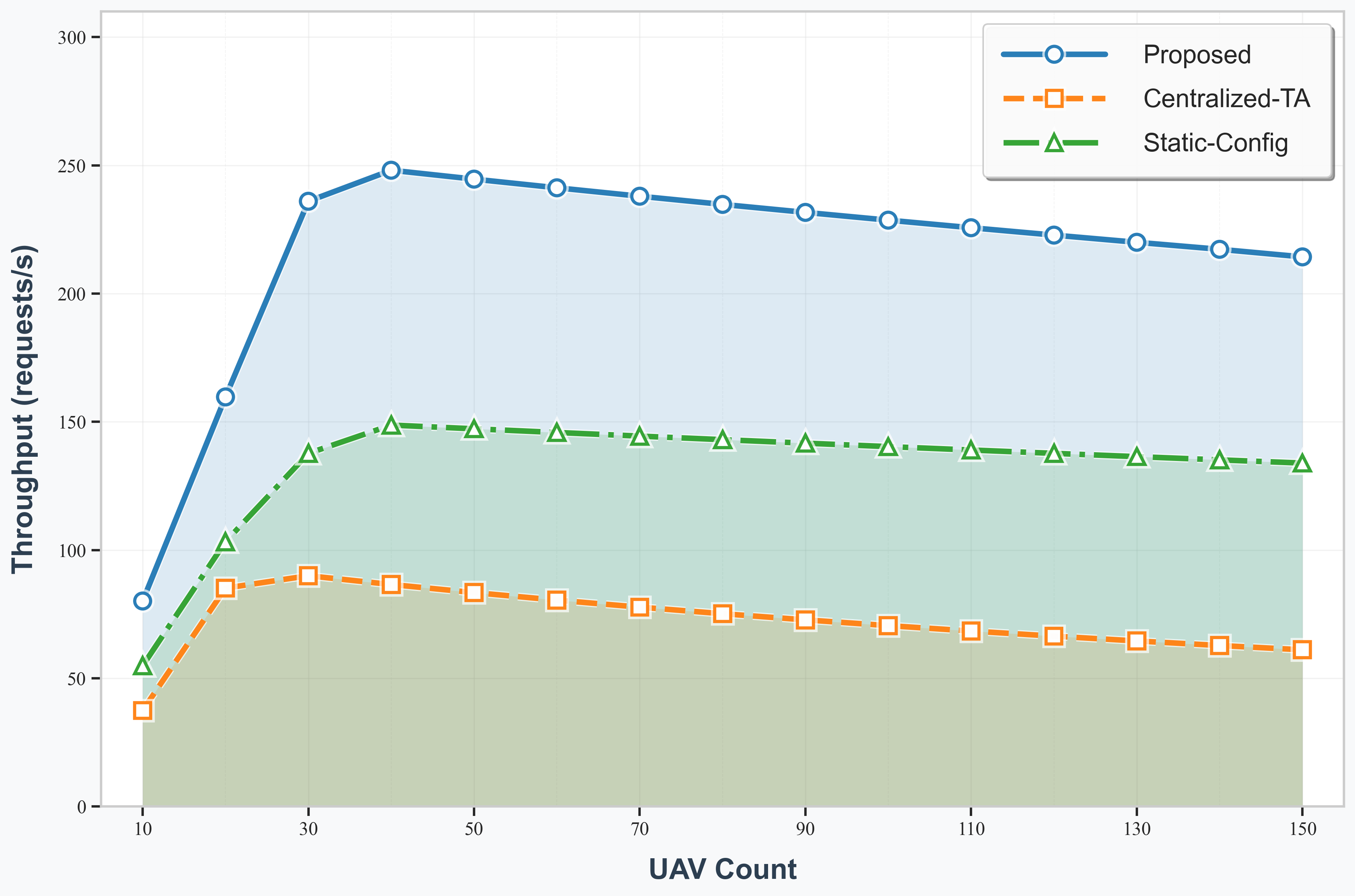}
    \caption{System throughput comparison under varying UAV count.}
    \label{fig:throughput}
\end{figure}

The system throughput denotes the number of SFC authentication tasks successfully completed within a given time interval. It reflects the capability of a system to support concurrent service requests. Fig.~\ref{fig:throughput} shows that the proposed method consistently achieves the highest throughput as the number of UAV nodes increases. The centralized-TA method performs poorly, because all authentication activities must be executed by a single central authority. This design reaches saturation quickly and forms a major bottleneck. The static-config method achieves moderate performance. Although the orchestrator is located at the network edge, the absence of dynamic selection causes imbalanced scheduling that constrains scalability. The proposed method provides superior throughput due to its distributed and adaptive structure. The dynamic election distributes orchestration responsibility among multiple edge servers, which prevents overload at a single coordination point. In addition, the streamlined authentication workflow reduces both computation cost and communication overhead during the cross-domain relay. These features enable the system to manage a large number of simultaneous SFC requests and demonstrate strong scalability in dense UAV environments.

The experimental results verify that the proposed blockchain-enabled trust framework significantly outperforms both the centralized-TA and static-config methods in authentication latency and system throughput. The evaluation confirms the advantage of combining distributed trust, dynamic orchestrator election, and lightweight cross-domain authentication. The proposed framework eliminates performance bottlenecks in traditional solutions while maintaining strong security guarantees. This framework therefore provides a practical foundation for scalable and resilient UAV collaborative networks.

\section{Challenges and Future Research Directions}

This section explores the key technical challenges in consortium blockchain-based trust frameworks for UAV networks and outlines potential future research directions to advance their development. A summary of the primary issues and corresponding research trends is provided in Table~\ref{tab:challenges_future}.

\subsection{Challenges}

The consortium blockchain-based cross-domain authentication frameworks hold significant promise for UAV networks, yet several technical hurdles must be overcome to enable large-scale practical deployment. A central challenge is ensuring scalability as the numbers of UAV nodes and transactions continue to increase. The expanding blockchain ledger can lead to increased storage and synchronization demands, particularly for edge servers and UAVs with constrained computational resources. Additionally, the high mobility and dynamic topology of UAV networks complicate timely trust evaluation and adaptive access control in rapidly evolving contexts, potentially causing discrepancies between the recorded trust states and real-time conditions.

Privacy protection is another critical challenge in consortium blockchain-enabled cross-domain UAV orchestration. Since the ledger records identity-related events and service handover traces across domains, long-term linkage of identifiers and correlated access to time-stamped handover records may allow inference of behavioral patterns and mission relationships. Therefore, privacy-aware logging and access governance are necessary to balance auditability and confidentiality in cross-domain deployments.

The interoperability and energy efficiency also pose notable challenges. The UAV systems across domains and manufacturers often utilize diverse communication standards and management policies, hindering seamless cross-domain identity verification and service coordination. Furthermore, the cryptographic operations and frequent message exchanges in blockchain and authentication processes can impose substantial energy demands on resource-limited UAVs. The need for lightweight cryptographic primitives and efficient consensus algorithms tailored to distributed edge environments remains critical. Addressing the above issues requires comprehensive optimization across network, computational, and energy dimensions.

\begin{table}[t] 
\centering
\caption{Challenges and Corresponding Research Directions}
\label{tab:challenges_future} 
\definecolor{tablehead}{RGB}{204, 229, 255} % 表头浅蓝
\begin{tabular}{>{\raggedright\arraybackslash\setlength{\parindent}{1em}}m{0.33\linewidth} 
                >{\raggedright\arraybackslash\setlength{\parindent}{1em}}m{0.58\linewidth}}
\toprule[1.2pt]
\rowcolor{tablehead}\textbf{Challenge} & \cellcolor{tablehead}\textbf{Research Direction} \\
\midrule
\multirow{3}{=}{Scalability and ledger growth} 
& - Lightweight blockchain architectures \\
& - Efficient off-chain data management \\
& - Reduced synchronization overhead \\
\midrule
\multirow{3}{=}{Dynamic trust and real-time security} 
& - Intelligent trust assessment \\
& - Adaptive authentication algorithms \\
& - Mobility-aware security monitoring \\
\midrule
\multirow{3}{=}{Privacy protection and data governance}
& - Dynamic pseudonyms \\
& - Off-chain logs with on-chain hash \\
& - Fine-grained query access control \\
\midrule
\multirow{3}{=}{Cross-domain interoperability} 
& - Standardized blockchain interfaces \\
& - Secure cross-chain operations \\
& - Unified identity management \\
\midrule
\multirow{3}{=}{Energy consumption in constrained devices} 
& - Low-power cryptographic primitives \\
& - Hardware acceleration \\
& - Communication overhead reduction \\
\bottomrule[1.2pt]
\end{tabular}
\end{table}  

\subsection{Future Research Directions}

The future research should prioritize advancements in scalability, adaptability, and efficiency for the blockchain-assisted UAV authentication systems. A key direction involves creating lightweight blockchain architectures with off-chain or sharded storage solutions to minimize the ledger size while preserving security and auditability. The adaptive trust management frameworks could incorporate artificial intelligence-based anomaly detection models for real-time, risk-aware authentication and policy adjustments. These approaches can enable UAV networks to respond dynamically to shifting operational environments and security landscapes, fostering a more resilient cross-domain trust infrastructure.

Moreover, efforts should be focused on the cross-domain standardization and energy optimization. Developing open interfaces and standardized protocols for consortium blockchain interoperability will facilitate the integration of heterogeneous UAV systems on a broader scale. For lightweight cryptography, elliptic-curve signatures and authenticated encryption can secure credentials with low computation and compact tokens. For energy efficiency, verification and auditing can be offloaded to edge servers, while UAV-side signaling is reduced through localized handover checks and on-chain hash logging. Exploring hardware-assisted acceleration and co-designed software-hardware authentication frameworks offers further potentials for enhancing the computational efficiency. In summary,       the future research aims to achieve an optimal balance of security, performance, and sustainability.

\section{Conclusions}
In this work, we have proposed a consortium blockchain-based trust framework for resilient cross-domain UAV SFC orchestration. Specifically, the framework has combined dynamic, service-aware orchestrator election with a four-phase cross-domain authentication protocol, thereby ensuring secure, efficient, and auditable collaboration among UAVs across multiple domains.  The case study results have demonstrated substantial improvements in authentication latency and system throughput compared to centralized and static orchestration schemes. Meanwhile, challenges and future directions related to scalability, adaptability, interoperability, and energy efficiency have been assessed to highlight key limitations and inform further research. Accordingly, future researches should prioritize lightweight consortium blockchain designs to reduce ledger overhead in our framework, together with adaptive trust management frameworks and energy-aware security solutions, to further enhance the performance and resilience of cross-domain UAV service orchestration.

\bibliographystyle{IEEEtran}

\bibliography{zeng2016wireless,yan2025multi,lu2024uav,wu2024adaptive,yang2023toward,wei2024hierarchical,feng2022blockchain,
tan2022blockchain,gong2023lcdma,chen2022xauth,pu2024litegap,hafeez2023blockchain,tan2022,zhou2019online,dolati2023layer,mao2025blockchain}

\end{document}